\documentclass[12pt]{article}
\usepackage{amsfonts,amsthm,amsmath,amssymb,graphicx,hyperref,youngtab}
\usepackage{cite}

\def\H{{\mathcal{H} }}
\def\dvf{{\overline{\delta\varphi} }}
\def\vf{{\varphi }}

\newcommand{\be}{\begin{equation}}
\newcommand{\bea}{\begin{eqnarray}}
\newcommand{\ee}{\end{equation}}
\newcommand{\eea}{\end{eqnarray}}
\newcommand{\nn}{\nonumber}

\begin{document}

\begin{center}

\begin{flushright}

{\small WITS-CTP-115
 \\
} \normalsize
\end{flushright}
\vspace{0.9 cm}

{\LARGE \bf{Cosmology or Catastrophe?\\ A non-minimally coupled scalar in an inhomogeneous universe\\\vspace{0.2cm}}}

\vspace{1.1 cm} {\large Pawe{\l} Caputa$^{a,b}$, Sheikh Shajidul Haque$^a$, Joseph Olson$^{c,d}$, and Bret Underwood$^{d}$}\,\footnote{{\ttfamily {pawel.caputa@wits.ac.za, shajid.haque@wits.ac.za, olson24@wisc.edu, bret.underwood@plu.edu }} }\\

\vspace{0.9 cm}
\vspace{0.2 cm}{{\it$^{a}$NITheP, School of Physics and Centre for Theoretical Physics,\\University of the Witwatersrand, Johannesburg, WITS 2050, South Africa }} \\
\vspace{0.2 cm}{{\it$^{b}$Tata Institute of Fundamental Research, Homi Bhabha Road, Mumbai 400005, India}}\\
\vspace{0.2 cm}{{\it$^{c}$Department of Physics, University of Wisconsin-Madison, Madison, WI 53706}}\\
\vspace{0.2 cm}{{\it$^{d}$Department of Physics, Pacific Lutheran University, Tacoma, WA 98447}}

\thispagestyle{empty}
\vspace{2cm}

{\bf Abstract}
\end{center}

\begin{quotation}

A non-minimally coupled scalar field can have, in principle, a negative effective Planck mass squared which depends on the scalar field.
Surprisingly, an isotropic and homogeneous cosmological universe with a non-minimally coupled scalar field
is perfectly smooth as the rolling scalar field causes the effective Planck mass to change sign and pass through zero.
However, we show that any small deviations from homogeneity diverge as the effective Planck mass vanishes, with catastrophic
consequences for the cosmology.
The physical origin of the divergence is due to the presence of non-zero scalar anisotropic stress
from the non-minimally coupled scalar field.
Thus, while the homogeneous and isotropic cosmology appears surprisingly sensible when the effective Planck mass
vanishes, inhomogeneities tell a different story.

\end{quotation}

\setcounter{page}{0}
\setcounter{tocdepth}{2}
\newpage


\section{Introduction}

The minimal coupling of a scalar field to gravity in curved space is through the curved metric replacement $\eta_{\mu\nu} \rightarrow g_{\mu\nu}$.
However, scalar fields can also couple directly to the scalar curvature, as in the non-minimally coupled action
\be
S=\int d^4x\sqrt{-g}\left(-\frac{m_p^2}{2}R+\frac{1}{2}g^{\mu\nu}\partial_{\mu}\varphi\partial_{\nu}\varphi-V(\varphi)+\frac{1}{2}\xi\, \varphi^2 R\right)\, .
\label{action0}
\ee
The non-minimal coupling parameter $\xi$ is dimensionless and can be any value; for example, 
when $\xi = 1/6$ the gravity-scalar system has a well-known conformal symmetry, while $\xi \gg 1$ has been used
in models of inflation \cite{Spokoiny:1984bd,Fakir:1990eg,Bezrukov:2007ep,Linde:2011nh} (although see \cite{Burgess:2010zq,Hertzberg:2010dc} for a discussion of problems with $\xi \gg 1$).
This type of direct coupling also appears in inflationary models as the {\it eta problem} \cite{Copeland:1994vg}; see also \cite{McAllister:2007bg} for a review.
The latter effect underscores a more general property of non-minimal coupling: even if $\xi = 0$ at tree level, it is expected
that $\xi$ will be generated by RG flow (see \cite{Einhorn:2009bh} and discussion within).

The physical effect of the coupling in (\ref{action0}) is more transparent after rearranging terms,
\be
S=\int d^4x\sqrt{-g}\left(-\frac{1}{2}\left(m_p^2-\xi\, \varphi^2\right)R+\frac{1}{2}g^{\mu\nu}\partial_{\mu}\varphi\partial_{\nu}\varphi-V(\varphi)\right)\, .\label{action}
\ee
Now the non-minimal coupling appears as a field-dependent ``{\it Effective Planck Mass}" $m_{p,eff}^2 = m_p^2 - \xi \varphi^2$.
For $\xi > 0$, there is a critical value of the scalar field $\vf_* \equiv m_p/\sqrt{\xi}$
such that at $\vf = \vf_*$ the effective Planck mass vanishes $m_{p,eff}^2 = 0$, and for $\vf> \vf_*$, the effective (squared) Planck mass is negative $m_{p,eff}^2 < 0$!

Certainly, there is reason to doubt that this region can actually be reached in the true quantum system; for example, as $m_{p,eff}^2 \rightarrow 0$, one would
expect that higher order Planck-suppressed operators like $R^2$ would become important, modifying the dynamics of the system considerably.
However, in this paper we wish to take the classical action (\ref{action}) at face value, and determine if there is something wrong with $m_{p,eff}^2 \leq 0$ {\it classically}.

In particular, we will consider (\ref{action}) on a cosmological background in which a homogeneous scalar field $\vf = \vf(t)$ begins above the critical value $\vf > \vf_*$,
so that $m_{p,eff}^2 < 0$ initially. The scalar field then rolls down potential so that $\vf(t)$ decreases with time; eventually $\vf = \vf_*$, and 
the effective Planck mass vanishes $m_{p,eff}^2 \rightarrow 0$.
Surprisingly, the homogeneous universe evolves smoothly through this critical point, as has been noted by other authors\footnote{We would like to thank Gary Felder
for conversations on this point.}\footnote{See also \cite{Maleknejad:2011sq,Maleknejad:2012as} for discussion in the context of the gauge-flation models} \cite{Linde:1979kf,starobinskii1981can,Futamase:1987ua} (note, however, that closed cosmological models do not evolve smoothly through this critical
point \cite{Morishima:1998ik}).
While a homogeneous and isotropic universe smoothly evolves through this critical point, deviations from homogeneity and isotropy would
be expected to cause an obstruction at vanishing effective Planck mass.
Indeed, earlier studies have shown that the presence of anisotropies cause a curvature singularity at the critical point\footnote{Perturbations
of black hole solutions of (\ref{action}) also have been shown to lead to instabilities \cite{Bronnikov:1978mx}.} \cite{starobinskii1981can,Futamase:1987ua,Futamase:1989hb}.

The purpose of this study is to explore the behavior of {\it inhomogeneous, but isotropic,} perturbations, in contrast to the previous studies,
as the effective Planck mass evolves through zero. In the process, we will derive the equations of cosmological perturbation theory \cite{Mukhanov1992203} for
the system (\ref{action}) in terms of the gauge-invariant variables. Note that since $m_{p,eff}^2$ is changing sign (and going through zero), it is not possible
to Weyl-rescale the cosmological solutions and study them in Einstein-frame, where much of the cosmology of non-minimally coupled scalar fields has been studied.

The paper is organized as follows. In Section 2, we review the homogeneous cosmological solutions of (\ref{action}), and demonstrate that the solutions
are perfectly smooth, even through the region where the effective Planck mass vanishes. In Section 3.1, we review cosmological perturbation theory
for a minimally coupled scalar field, and show that if $m_p^2 \rightarrow 0$ for minimal coupling, then all of the perturbations must be trivial.
In Section 3.2 we derive the equations of cosmological perturbation theory for (\ref{action}), concentrating on the extra terms arising from the non-minimal coupling.
We will show there that as the homogeneous background evolves through the critical point, the perturbations must either be trivial or divergent, signaling a
cosmological ``catastrophe." 
In Section 4, we conclude with a brief discussion. Some detailed calculations are relegated to the appendices.

\section{Cosmology through the Critical Point}

First, we will explore the homogeneous cosmology of the non-minimally coupled scalar field as it passes through the critical point.
We will show that, surprisingly, the cosmology is completely smooth and non-trivial through the critical point where the effective Planck mass vanishes due the non-minimal
coupling.

Starting from the action (\ref{action}) for the non-minimally coupled scalar field, Einstein's equations\footnote{In principle, it is necessary to include a generalization of the Gibbons-Hawking-York
term $S_{GHY} = \int_{\partial M} \sqrt{h} (m_p^2-\xi\vf^2) K$ following \cite{Guarnizo:2010xr}, where $h$ is the induced metric on the boundary $\partial M$ of the manifold, and $K$ is the trace of the extrinsic curvature,
in order to cancel the boundary terms generated by variation of (\ref{action}).}
and the scalar field equation of motion take the form:
\bea
&&(m^2_p-\xi\vf^2)G^{\mu}_{\nu}=g^{\mu\alpha}\vf_{,\alpha}\vf_{,\nu}-\delta^{\mu}_{\nu}\left(\frac{1}{2}g^{\alpha\beta}\vf_{,\alpha}\vf_{,\beta}-V(\vf)\right)+\xi\left(\delta^{\mu}_{\nu}\,\Delta\vf^2-\nabla_{\nu}\nabla^{\mu}\vf^2\right)\, ; \label{EENonMin} \\
&& \frac{1}{\sqrt{-g}}\partial_{\mu}\left(\sqrt{-g}g^{\mu\alpha}\partial_{\alpha}\varphi\right)+\frac{\partial V(\varphi)}{\partial\varphi}=\xi R\,\vf\, . \label{NMScalarEOM}
\eea
Notice that there are two places that a non-zero non-minimal coupling $\xi$ leads to additional terms: First, in the coefficient of the Einstein tensor, modifying the effective Planck mass. 
Second, as additional ``energy-momentum" terms acting as sources on the right hand side of Einstein's equation\footnote{As pointed out in \cite{0264-9381-5-4-010} these extra contributions create an ambiguity in the definition of the energy-momentum tensor $T^\mu_\nu$. 
If $T^\mu_\nu$ is defined formally as the functional derivative of the matter action $T_{\mu\nu} = \frac{2}{\sqrt{g}}\frac{\delta {\mathcal L}_m}{\delta g^{\mu\nu}}$, it will not be divergence free. 
However, if we define $T^\mu_\nu$ to be equal to the source of the Einstein tensor in the Einstein equations, $m_p^2 G^\mu_\nu = T^\mu_\nu$, then the divergence will vanish but it will not be the same as that obtained by variation of the matter action.
Our organization of the Einstein equations (\ref{EENonMin}) reflects the structure of the latter approach, although the two definitions are just related by algebra.}.

\begin{figure}[t]
\centering \includegraphics[height=.5\textwidth]{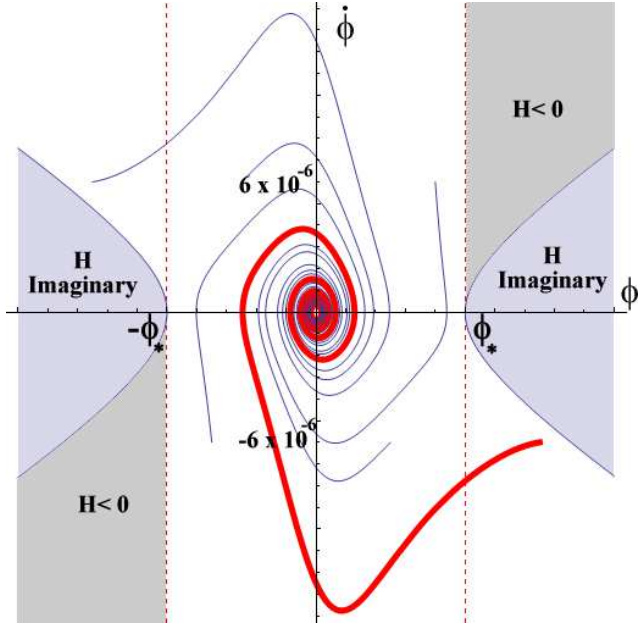}
\caption{The phase space for the system (\ref{FriedmanNonMin},\ref{BackgroundScalarEOM}), for $V(\vf) = \frac{1}{2} m^2 \vf^2$, with $m = 10^{-5} m_p$ and $\xi = 1$, in units where $m_p = 1$. The regions
of imaginary $H$ and negative $H$ are separated from the positive $H$ region by infinite $H$, so that it is not possible to flow from one such region of phase space to another. Notice
that the highlighted trajectory smoothly passes through the critical point $\vf = \vf_*$ at which the effective Planck mass vanishes.}
\label{phasespace}
\end{figure}

\begin{figure}[t]
\centering \includegraphics[width=.6\textwidth]{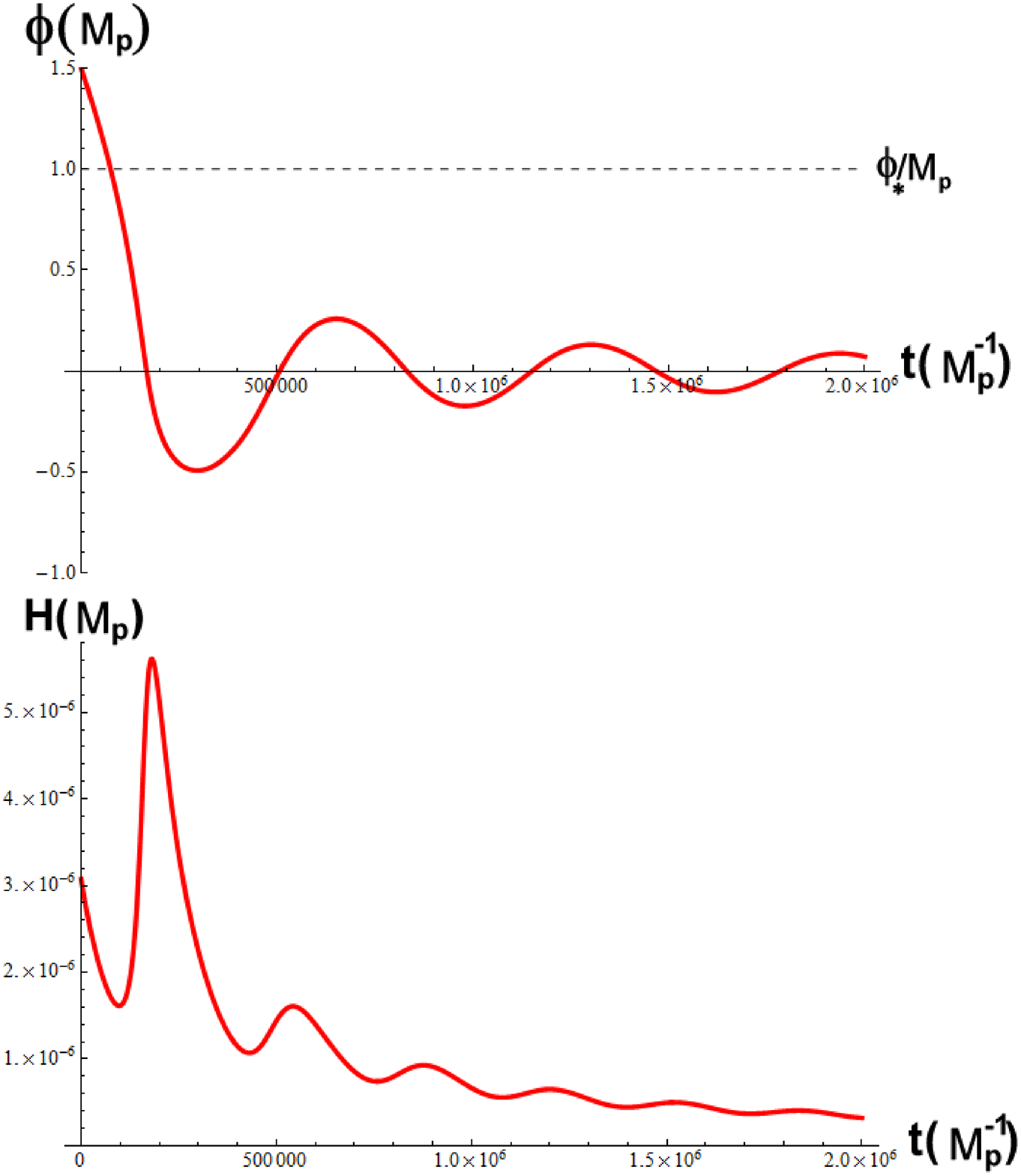}
\caption{Top: The evolution of the scalar field $\vf(t)$ versus time for the highlighted trajectory from Figure \ref{phasespace} is shown. Note that the evolution is completely smooth through the
critical point $\vf_* = m_p/\sqrt{\xi}$. Bottom: The evolution of the Hubble parameter $H(t)$ versus time for the same trajectory. Note
that the Hubble parameter $H(t)$ is finite and smooth as the scalar field passes through the critical point.}
\label{ScalarvsTimePlot}
\end{figure}

Assuming a homogeneous and isotropic (and flat) universe, the metric and scalar field take the form,
\bea
ds^2 &=& dt^2 - a(t)^2 d\vec{x}^2\,;  \\
\varphi &=& \varphi_0(t)\, .
\eea
Inserting these into the Einstein and scalar field equations above, we obtain the equations for the background scale factor and scalar field $a(t),\varphi(t)$
\bea
&& 3(m_p^2-\xi \varphi_0(t)^2) H^2 = \frac{1}{2} \dot \varphi_0^2 + V(\varphi_0)+6\, \xi\, \varphi_0(t) \dot\varphi_0\, H\, ; \label{FriedmanNonMin} \\
&& \ddot\varphi_0 + 3 H \dot \varphi_0 + V'(\varphi_0) + 6\, \xi(2 H^2 + \dot H) \vf_0= 0\, . \label{BackgroundScalarEOM}
\eea
where $H(t) = \frac{\dot a}{a}$ is the Hubble expansion parameter. 

Remarkably, it is possible to have a solution to the modified Friedmann equation (\ref{FriedmanNonMin}) even when the effective Planck mass vanishes because of
the additional term arising from the non-minimal coupling. In particular, the general solution of (\ref{FriedmanNonMin}) for the Hubble parameter $H$ in an expanding universe is
\be
H = \frac{6\xi\varphi_0 \dot\varphi_0 + \sqrt{(6 \xi \varphi_0 \dot \varphi_0)^2 + 12 (m_p^2-\xi \varphi_0^2) (\frac{1}{2} \dot\varphi_0^2 + V(\varphi_0))}}{6 (m_p^2 - \xi \varphi_0^2)}\, .
\ee
When the scalar field is close to the critical point $\varphi_0\approx \vf_* = m_p/\sqrt{\xi}$ , the Hubble parameter becomes,
\be
H \approx \frac{\xi (\varphi_0\dot\varphi_0+|\varphi_0\dot\varphi_0|)}{m_p^2-\xi\varphi_0^2}+\frac{\frac{1}{2}\dot \varphi_0^2 +V(\varphi_0)}{6\xi |\varphi_0 \dot\varphi_0|}\, .
\ee
When $\varphi_0$ or $\dot\varphi_0$ are separately negative, the first term vanishes and the Hubble parameter is finite even when the effective Planck mass vanishes.
If, however, the product $\varphi_0 \dot \varphi_0$ is positive, the first term does not vanish and the Hubble parameter becomes infinite when the effective Planck mass vanishes.
Similarly, in regions where the effective Planck mass is negative, real solutions only exist when $\xi \varphi_0\dot\varphi_0$ is sufficiently large, otherwise the Hubble parameter is
imaginary, signaling the existence of AdS solutions instead of cosmological solutions. This structure is shown in the phase-space plot of Figure \ref{phasespace}.

Remarkably, it is possible to start in a region where the effective Planck mass is negative with real and positive Hubble parameter, and smoothly evolve through the critical point, as 
seen by the highlighted trajectory in Figure \ref{phasespace}. Figure \ref{ScalarvsTimePlot} displays the value of the scalar field for this trajectory
as a function of time. Notice again that the system smoothly passes through the critical point $\vf=\vf_*$ where the effective Planck mass vanishes.
Thus, the cosmology of the non-minimal scalar field is smooth and continuous, even through the critical point!

\section{Catastrophe - Perturbations through the Critical Point}

\subsection{Perturbations for Minimally Coupled Scalar Field}

Let us review cosmological perturbation theory for a minimally coupled ($\xi = 0$) scalar field $\varphi$, as can be found in \cite{Mukhanov1992203}.
In longitudinal gauge, the metric and scalar perturbations are
\bea
&&ds^2=a^2\left[(1+2\phi)d\eta^2-(1-2\psi)\delta_{ij} \ dx^idx^j\right];\label{metscalar} \\
&&\vf=\varphi_{0}(\eta)+\delta\varphi(x,\eta).\label{flucscalar}
\eea
Note that we have switched to conformal time $\eta$, defined by $\eta = \int a(t)^{-1} dt$, which is a more convenient time variable for cosmological
perturbations.
The Einstein equations, to first order in the perturbations, are then,
\begin{eqnarray}
\label{MinimalE1} a^2m^{2}_p\, \overline{\delta G^0_0} &=&  2\,m^{2}_p\left[\Delta\Psi-3\H\Psi'-(\H'+2\H^2)\Phi\right]=\vf'_0\,\dvf'+\dvf\,a^2\,V_{,\vf}\,;\\
\label{MinimalE2}a^2m^{2}_p\,  \overline{\delta G^0_i} &=& 2\,m^{2}_p(\Psi'+\H\Phi)_{,i}=(\vf'_0\,\dvf)_{,i}\,;\\
a^2m^{2}_p\, \overline{\delta G^i_j} &=& 2\,m^{2}_p\left[\Psi''+\H (2\Psi+\Phi)'+(\H'+2\H^2)\Phi+\frac{1}{2}\Delta(\Phi-\Psi)\right]\delta^i_j \nonumber \\
\label{MinimalE3}&&- m^{2}_p(\Phi-\Psi)_,\, {}^i_j = \left[\vf'_0\,\dvf'-\dvf\,a^2\,V_{,\vf}\right]\delta^i_j\, ;
\end{eqnarray}
where we have replaced the perturbations with their corresponding gauge-invariant variables $\phi\rightarrow \Phi, \psi \rightarrow \Psi, \delta\varphi \rightarrow \dvf$
(see \cite{Mukhanov1992203} for more details on the gauge-invariant variables $\Phi,\Psi,\dvf$).

Notice that the off-diagonal terms of (\ref{MinimalE3}) require $\Phi = \Psi$, implying the absence of anisotropic stress due to the perturbations of the scalar field. 
As we will see later, the surprising presence of a source of anisotropic stress for the non-minimal scalar field
leads to some significant consequences for the behavior of the perturbations.

In order to develop some physical intuition, let's consider the equations (\ref{MinimalE1}-\ref{MinimalE3})
in the limit where $m_p^2 \rightarrow 0$. Certainly, the non-minimal coupling introduces many additional
terms, so the non-minimal coupling is {\it more} than just controlling the size and sign of $m_p^2$. However,
analyzing this limit will allow us to learn about how the equations behave under this limit, and what is ``normal"
and what is due to the non-minimal coupling.

In the limit $m_p^2\rightarrow 0$, the equations (\ref{MinimalE1}-\ref{MinimalE3}) (as well as the minimal version of the background equation of motion (\ref{FriedmanNonMin})
and the perturbed version of the scalar field equation of motion (\ref{NMScalarEOM})) become
\begin{eqnarray}
0 & = & \dvf''+2\H\dvf'-\Delta \dvf + a^2V_{,\vf\vf}\dvf - \vf_0'(3\Psi+\Phi)'+2a^2V_{,\vf} \Phi = 0\,; \\
0 & = & \vf_0 \dvf+ \dvf a^2 V_{,\vf}\,; \\
0 & = & \left(\vf_0' \dvf\right)_{,i}\,; \\
0 & = & \vf_0'\dvf' - \dvf a^2 V_{,\vf}\,; \\
0 & = & \vf_0'^2\,.
\end{eqnarray}
These equations require trivial solutions
\begin{eqnarray}
\dvf \rightarrow  0\,; \hspace{.3in} \vf_0' \rightarrow 0\,; \hspace{.3in}\Psi \rightarrow  0\,; \hspace{.3in} \Phi \rightarrow 0\,;
\end{eqnarray}
so that for a minimally coupled scalar field, a vanishing Planck mass only allows trivial solutions.

\subsection{Perturbations for Non-Minimally Coupled Scalar Field}

For the non-minimally coupled scalar field, the Einstein equations (\ref{EENonMin}) in longitudinal gauge (\ref{metscalar},\ref{flucscalar}) become at linear 
order\footnote{Again, we have replaced $\phi, \psi, \varphi$ with their corresponding gauge-invariant quantities $\Phi,\Psi,\dvf$, which are defined in the same
way as for minimal coupling.}
\begin{eqnarray}
\label{FullPert1}(m^2_p-\xi\vf^2_0)a^{2}\delta G^{0}_{0}-6\xi\H^2\vf_0\dvf&=&-\Phi\,(\vf'_0)^2+\vf'_0\dvf'+\dvf\,a^2 V_{,\vf}\\
                                                                                                      &-&2\xi \left(\vf_0\Delta\dvf-3\H(\vf_0\dvf)'+3\vf_0\vf'_0(\Psi'+2\H\Phi)\right)\,;\nn\\
(m^2_p-\xi\vf^2_0)a^{2}\delta G^{0}_{i}&=&\vf'_0\,\dvf_{,i}-2\xi\left((\vf_0\,\dvf)'-\vf_0(\H\dvf+\Phi\,\vf'_0)\right)_{,i}\,;\nn\\
\label{FullPert2}\\
(m^2_p-\xi\vf^2_0)a^{2}\delta G^{i}_{j}-2\xi(\H^2+2\H')\vf_0\,\dvf\,\delta^{i}_{j}&=&2\xi\,\vf_0\,\dvf_{,ij}+\delta^{i}_{j}\left(\Phi\,(\vf'_0)^2-\vf'_0\,\dvf'+\dvf\,a^2V_{,\vf}\right.\nn\\
&-&\left.2\xi\left(\vf_0\Delta\dvf+2\Phi(\H\vf_0\vf'_0+(\vf'_0)^2+\vf_0\vf''_0)\right.\right.\nn\\
&+&\left.\left.(\Phi+2\Psi)'\vf_0\vf'_0-(\vf_0\,\dvf)''-\H(\vf_0\,\dvf)'\right)\right)\,;
\label{FullPert3}
\end{eqnarray}
with the $\delta G^{\mu}_{\nu}$ given by (\ref{MinimalE1}-\ref{MinimalE3}).
Consider (\ref{FullPert3}) for $i\neq j$:
\begin{eqnarray}
\label{FullPert4}( m^2_p-\xi\vf^2_0)(\Phi-\Psi)_{,ij}=-2\xi(\vf_0\,\dvf)_{,ij}\, .
\end{eqnarray}
Notice that the non-minimal coupling introduces non-zero scalar anisotropic stress.
Normally, scalar anisotropic stress is generated by quadrupole radiation fields.
For instance, in the evolution of scalar perturbations in our Universe, the only appreciable
amount of scalar anisotropic stress comes from the quadrupole moment of the
neutrino background radiation \cite{Dodelson}.

Let's now explore the behavior of the equations as the background passes through the critical point, where the effective Planck mass vanishes. 
Since the homogeneous background is smooth and continuous through the critical point, the vanishing of the effective Planck mass only occurs at a specific instant of time $\eta^*$, where $\vf_0(\eta^*) = \vf_* = m_p/\sqrt{\xi}$.
Further, because the equations are linear we can study the behavior of individual Fourier modes of the perturbations
\begin{equation} \label{Fourier}
\dvf(x,\eta) = \dvf_k(\eta) e^{i\vec{k}\cdot \vec{x}},\hspace{.5in} \Phi(x,\eta) = \Phi_k(\eta) e^{i\vec{k}\cdot \vec{x}},\hspace{.5in} \Psi(x,\eta) = \Psi_k(\eta) e^{i\vec{k}\cdot \vec{x}};
\end{equation}
since different Fourier modes decouple.
The off-diagonal Einstein equation (\ref{FullPert4}) then becomes
\be
\label{NonMinimalE3v2}(m^2_p-\xi\vf^2_0)(\Phi_k-\Psi_k)=-2\xi\vf_0\,\dvf_k\, .
\ee
As the critical point is approached, the coefficient of the left-hand-side of (\ref{NonMinimalE3v2}) vanishes, but the right-hand-side does not
necessarily vanish.
Thus, if $\dvf_k(\eta^*) \neq 0$ at the critical point, then (\ref{NonMinimalE3v2}) requires that the perturbations diverge $\Phi_k,\Psi_k\rightarrow \infty$,
and thus we have a ``{\it catastrophe}" at the critical point!
As we argue in the Appendix, the vanishing of the scalar field fluctuation at the critical point $\dvf_k(\eta^*) = 0$ can only occur if all of
the fluctuations are trivial for all time. Thus, any non-zero inhomogeneous (but isotropic) perturbations to the smooth cosmological backgrounds discussed
in Section 2 lead to a ``{\it catastrophe}" as the effective Planck mass vanishes.

\section{Discussion}

We have seen that the homogeneous and isotropic cosmology of a non-minimally coupled
scalar field can smoothly evolve from negative effective Planck mass squared $m_{p,eff}^2 < 0$,
through the ``critical point" where $m_{p,eff}^2 = 0$, into the region where $m_{p,eff}^2 > 0$.
The absence of any divergence near the critical point is due to the presence of additional ``energy-momentum"
source terms in the Einstein equations. 

However, we have also shown that the presence
of these additional terms cause small isotropic inhomogeneous perturbations to this cosmological background to diverge at the critical point. Thus, while the homogeneous and isotropic cosmology appears to be sensible as it evolves through the critical point, any perturbation away from homogeneity will cause a divergence at the critical point.
This result is complementary to earlier results \cite{starobinskii1981can,Futamase:1987ua,Futamase:1989hb}
that found that any deviation from {\it isotropy} causes a divergence at the critical point.
Thus, we have found that it is not just deviations from isotropy, but also deviations from homogeneity,
that prevents a cosmological universe from passing through the critical point.
This classical obstruction to $m_{p,eff}^2 = 0$ is in addition to any additional quantum
effects that are expected to be important in this regime.

As discussed in Section 3.2, the origin of the divergence in the inhomogeneities is easy to see.
Of the additional energy-momentum source terms, there is a contribution to the 
scalar anisotropic stress that is non-vanishing at the critical point.
Recall that scalar anisotropic stress refers to the difference between the Newtonian potential
$\Phi$ and the perturbation to the spatial curvature $\Psi$. In the absence of anisotropic
stress, these two potentials are equal.
Normally, scalar anisotropic stress comes from the quadrupole moment of a cosmic fluid;
for a minimally coupled scalar field, the scalar anisotropic stress vanishes at linear order
in the perturbations. Its presence here for the non-minimally coupled scalar field is a surprise.
Since the scalar anisotropic stress is non-zero, even at the critical point, the gravitational
potentials $\Phi,\Psi$ must diverge there.

Of course, our analysis only studies the behavior of linear perturbations; it would
be interesting to study the full non-linear behavior in future work.
It would also be interesting to study how these additional energy-momentum terms modify
the usual cosmological perturbation theory and spectrum of inflationary perturbations.
We hope to explore this in the future.

\subsection*{Acknowledgements} 
We would like to thank Gary Felder for some initial discussion on this topic, and
Ilies Messamah for useful explanations and discussions on this topic. We also thank Rakibur Rahman and Bogomil Gerganov for useful discussions. The work of P.C.~and S.S.H.~is supported by the South African Research Chairs Initiative of the Department of Science and Technology and National Research Foundation. J.O.~and B.U.~would like to thank the Pacific Lutheran University Division of Natural Sciences for their support while this research was conducted. 
\begin{appendix}

\section{Trivial Solution} \label{eta issue}

In Section 3.2, our argument that inhomogeneities diverge when the background effective Planck mass vanishes
was based on $\dvf_k(\eta^*) \neq 0$ at the critical point. But it is possible that the evolution of $\dvf_k(\eta)$
be such that $\dvf_k(\eta^*) = 0$ precisely at the critical point, avoiding the ``catastrophe."
In this appendix, we argue that this is not the case - the only way that $\dvf_k(\eta^*) = 0$ is if
all of the fluctuations vanish for all time, thus implying that the only non-trivial solution at the critical point
is the ``catastrophe."

The equations (\ref{FullPert1}-\ref{FullPert4}) at the critical point become a set of equations for $\dvf_k(\eta^*),\Phi_k(\eta^*)$ and $\Psi_k(\eta^*)$ (where
we drop the functional dependence on $\eta^*$ for notational simplicity):
\begin{eqnarray}
6\xi\H^2\vf_*\dvf_k-\Phi_k\,(\vf'_*)^2-\vf'_*\dvf_k'-\dvf_k\,a^2 V_{,\vf}-2\xi \left[-k^2\vf_*\dvf_k-3\H(\vf_*\dvf_k)' \right.&& \nn \\
\label{NonMinimalE1}\left.+3\vf_*\vf'_*(\Psi_k'+2\H\Phi_k)\right]= 0\, ; && \hspace{.2in} \\
\label{NonMinimalE2}k_i \left[\vf'_*\,\dvf_k-2\xi\left((\vf_*\,\dvf_k)'-\vf_*(\H\dvf_k+\Phi_k\,\vf'_*)\right)\right] =0\, ;  && \\
\label{NonMinimalE3}k_i k_j 2\xi \vf_*\,\dvf_k = 0\, ; && \\
2\xi(\H^2+2\H')\vf_*\,\dvf_k- \vf'_*\,\dvf_k'+\Phi_k\,(\vf'_*)^2+\dvf_k\,a^2V_{,\vf}+ \frac{2}{3} k^2\xi \vf_*\dvf_k  &&\nn\\
-2\xi\left[-k^2\vf_*\dvf_k-(\vf_*\,\dvf_k)''-\H(\vf_*\,\dvf_k)'\right.&& \nn \\
\label{NonMinimalE4}\left.+2\Phi_k(\H\vf_*\vf'_*+(\vf'_*)^2+\vf_*\vf''_*)+(\Phi_k+2\Psi_k)'\vf_*\vf'_*\right] = 0\, . &&
\end{eqnarray}

Clearly, (\ref{NonMinimalE3}) is solved by $\dvf_k(\eta^*) = 0$. Combining and simplifying (\ref{NonMinimalE1},\ref{NonMinimalE2}) and (\ref{NonMinimalE4}),
we obtain the three simplified equations:
\bea
&& \dvf_k' - \vf_0' \Phi_k = 0\, ; \label{TimeSpaceConstraint} \\
&& \Psi_k'+ \H \Phi_k = 0\,; \label{TimeTimeConstraint} \\
&& \vf_0 \left(\Phi_k \vf_0' - \dvf_k'\right)' +\left(\frac{\vf_0\vf_0''}{\vf_0'}- \H \vf_0\right)\dvf_k' = 0\, .\label{delvpahiprime}
\eea
Using (\ref{FullPert2}), we can write
\be 
\left(\,\dvf'_k-\Phi_k\,\vf'_0\right)'= \left( -\frac{\vf'_0}{2 \xi \vf_0} -\frac{\vf'_0}{\vf_0}+\H\right ) \dvf'_k\, . \label{newconstraint}
\ee
Inserting this into (\ref{delvpahiprime}), we obtain
\be
\left[\left(1+\frac{1}{2 \xi } \right)\ \vf'_0 +\frac{\vf_0 \vf''_0}{\vf'_0} - 2 \H \vf_0\right ] \dvf'_k=0\, .
\ee
This can solved by either $\dvf_k'(\eta^*) = 0$ or with the term in parentheses vanishing. In general, the term in parentheses does
not vanish, but it could be possible to construct a very special solution that does so, similar to the special anisotropic cases found in
\cite{Futamase:1989hb}. Ignoring this very special solution, we will take
the more general case $\dvf_k'(\eta^*) = 0$.
Inserting $\dvf_k' = 0$ back into (\ref{TimeSpaceConstraint}-\ref{delvpahiprime}), we find that if $\dvf_k(\eta^*) = 0$, then not only
do the other perturbations vanish there as well $\Phi_k(\eta^*) = 0 = \Psi_k(\eta^*)$, but derivatives of the fluctuations also vanish:
\bea 
&&\dvf_k(\eta^*) = 0, \hspace{.3in} \dvf_k'(\eta^*) = 0, \hspace{.3in} \dvf_k''(\eta^*) = 0, \nn \\
&& \Phi_k(\eta^*) = 0, \hspace{.3in} \Phi_k'(\eta^*) = 0, \nn \\
&& \Psi_k(\eta^*) = 0, \hspace{.3in} \Psi_k'(\eta^*) = 0\, . \label{conclusion}
\eea

Since the equations of motion for $\dvf_k(\eta),\Psi_k(\eta),\Phi_k(\eta)$ are linear first and second order differential equations,
then the vanishing of the perturbations and their derivatives at the critical point implies that the perturbations themselves vanish for all time $\dvf_k(\eta) = 0 = \Psi_k(\eta) = \Phi_k(\eta)$.
In particular, the general form of the equations of motion for $\dvf_k,\Phi_k,\Psi_k$ are three independent equations of motion
\bea
\dvf_k'' + A_1(\eta) \dvf_k' + A_2(\eta) \dvf_k + A_3(\eta) \Psi_k' + A_4(\eta) \Psi_k + A_5(\eta) \Phi_k' + A_6(\eta) \Phi_k &=& 0\,; \label{eom1}\\
\Phi_k' + B_1(\eta) \Phi_k + B_2(\eta) \dvf_k'+B_3(\eta) \dvf_k + B_4(\eta) \Psi_k &=& 0 \,;\label{eom2}\\
C_0(\eta) \Psi_k''+C_1(\eta) \Psi_k' + C_2(\eta) \Psi_k + C_3(\eta) \dvf_k' + C_4(\eta) \dvf_k + C_5(\eta) \Phi_k &=& 0\,; \label{eom3}
\eea
where the $A_i(\eta), B_i(\eta), C_i(\eta)$ are some generally non-vanishing functions of $\eta$, except for $C_0(\eta)$, which we can see from (\ref{MinimalE3},\ref{FullPert3})
is proportional to $m_{p,eff}^2$, which vanishes at $\eta = \eta^*$.
These equations are all solved at $\eta = \eta^*$ by (\ref{conclusion}).
Taking a derivative of (\ref{eom1}-\ref{eom3}), we obtain, equations of the form,
\bea
\dvf_k''' + (A_1(\eta) \dvf_k')' + (A_2(\eta) \dvf_k)' + (A_3(\eta) \Psi_k')' + (A_4(\eta) \Psi_k)' + (A_5(\eta) \Phi_k')' + (A_6(\eta) \Phi_k)' &=& 0\,; \nn\\
&& \label{deom1}\\
\Phi_k'' + (B_1(\eta) \Phi_k)' + (B_2(\eta) \dvf_k')'+(B_3(\eta) \dvf_k)' + (B_4(\eta) \Psi_k)' &=& 0\,; \nn \\
&& \label{deom2}\\
(C_0(\eta) \Psi_k'')' + (C_1(\eta)\Psi_k')' + (C_2(\eta) \Psi_k)' + (C_3(\eta) \dvf_k')' + (C_4(\eta) \dvf_k)' + (C_5(\eta) \Phi_k)' &=& 0\,. \nn \\
&&\label{deom3}
\eea
Evaluating (\ref{deom1}-\ref{deom3}) at $\eta = \eta^*$ by imposing (\ref{conclusion}), we then find that the next order in derivatives must vanish,
\be
\dvf_k'''(\eta^*) = 0 = \Phi_k''(\eta^*) = \Psi_k''(\eta^*)\, .
\ee
Extending this argument to an arbitrary number of derivatives shows that {\it all} derivatives of $\dvf_k,\Phi_k,\Psi_k$ must vanish at the critical point.
If all derivatives vanish at $\eta = \eta^*$, then the perturbations themselves must vanish for all time $\dvf_k(\eta) = 0 = \Psi_k(\eta) = \Phi_k(\eta)$.
Thus, if $\dvf_k(\eta^*) = 0$, then all of the perturbations must be trivial.

\end{appendix}

\bibliography{refs}

\bibliographystyle{utphysmodb}

\end{document}